\documentclass[journal]{IEEEtran}
\usepackage{url}
\usepackage{algpseudocode}
\usepackage{algorithm}
\usepackage{graphicx}
\usepackage{caption}
\usepackage[labelformat=simple]{subcaption}

\usepackage{listings}
\usepackage{mathtools}
\usepackage{multirow}
\usepackage{color}
\usepackage{textcomp}

\newcommand{\ignore}[1]{}

\begin{document}

\title{APEX: Automatic Event Sequence Generation for Android Applications}
\author{%
  Wenhao Chen\IEEEauthorrefmark{1}, Morris Chang\IEEEauthorrefmark{2}, Witawas Srisa-an\IEEEauthorrefmark{3}, Yong Guan\IEEEauthorrefmark{4}\\
  \IEEEauthorrefmark{1} OpenText, {wenhaocisu@gmail.com}\\
  \IEEEauthorrefmark{2} University of South Florida, {chang5@usf.edu}\\
  \IEEEauthorrefmark{3} University of Nebraska at Lincoln, {witty@cse.unl.edu}\\
  \IEEEauthorrefmark{3} Iowa State University, {guan@iastate.edu}%
}

\maketitle

\begin{abstract}

Due to the event
driven nature and the versatility of GUI designs in Android
programs, it is challenging to generate event sequences with
adequate code coverage within a reasonable time. A common
approach to handle this issue is to rely on GUI models to generate
event sequences. These sequences can be
effective in covering GUI states, but inconsistent in exposing
program behaviors that require specific inputs. A major obstacle
to generate such specific inputs is the lack of a systematic GUI exploration process
to accommodate the analysis requirements.
In this paper, we introduce Android Path Explorer (APEX), a
systematic input generation framework using concolic execution.
APEX addresses the limitations of model-based sequence
generation by using concolic execution to discover the data dependencies
of GUI state transitions. Moreover, concolic execution
is also used to prioritize events during the exploration of GUI,
which leads to a more robust model and accurate
input generation. The key novelty of APEX is that
concolic execution is not only used to construct event sequences,
but also used to traverse the GUI more systematically.
As such, our experimental results show that APEX can be used to
generate a set of event sequences that achieve high code coverage,
as well as event sequences that reach specific targets.

\end{abstract}


\section{Introduction}


As mobile devices become increasingly prevalent, we
have seen a significant growth of mobile application
ecosystem in recent years. As the most widely used
mobile operating system~\cite{idc2017q1}, Android has
provided users with millions of apps in a variety of
categories.  In order to attract users in the highly
open and competitive Android app marketplace, app
developers need to deliver dependable programs that
also perform as described, and app store auditors need
to identify and remove any malicious apps from the
marketplace.  As a result, a great deal of research
have been conducted in the area of Android app testing.

%

Input generation is an important and challenging
technique used in program
testing~\cite{VisserISSTA04}~\cite{ChoudharyASE15}. Recently,
it also becomes quite important in dynamic program
analysis as modern Android applications increasingly
rely more on dynamic code loading and
reflection~\cite{StaDyna15}~\cite{PoeplauNDSS14} to perform
tasks, update components, and provide backward
compatibility with older devices and
platforms. When code are loaded at
runtime, static analysis is not capable of analyzing
such code as it is not available until runtime. Thus,
dynamic analysis is needed but the its
ability to produce quality results greatly depends on
the quality of the provided inputs.  Specifically, the
provided inputs must be able to reach code sections in
a program that can dynamically load
code~\cite{BoddenICSE11}. Otherwise, dynamic analysis
would be ineffective. 

In general, the goal of input generation
is to find a set of inputs that can trigger different
behaviors of the program, enabling further analysis on
those behaviors. Android apps are event-driven applications in
which the execution of the program is determined by
events such as user actions (tapping, swiping, etc.) or
system events (battery state change, incoming SMS, etc.).
As such, an input for an Android application
represents a sequence of events. Generating event
sequences with adequate code coverage while avoiding
potential explosion in the number of events sequences
is one of the main challenges of event sequence
generation.

Researchers have proposed many tools and algorithms
aiming to improve the effectiveness and efficiency of
input generation processes. Existing tools such as
monkey~\cite{monkey}, DynoDroid~\cite{machiry2013_dynodroid},
sapienz~\cite{sapienz} use random exploration
strategies to generate event sequences. These tools
usually treat an app as a black-box. For example,
\emph{Monkey} generates each event randomly without
inspecting the GUI, while DynoDroid extracts relevant
events from the GUI before selecting an event with a
biased random strategy. Random input generation
tools have the ability to generate a large amount of
events in a short amount of time, by avoiding expensive
operations such as code analysis and GUI state
inspection. However, their lack of precise control
over the exploration often results in inadequate code
coverage, especially in apps that have complex GUI
structures.

Model-based input generation tools such as
A\textsuperscript{3}E~\cite{azim2013_a3e},
MobiGUITAR~\cite{amalfitano2015_mobiguitar},
SwiftHand~\cite{choi2013_swifthand},
etc., can generate event sequences with better code
coverage by first building GUI models and then
generating event sequences based on these models.
Although each of the aforementioned tools has a
different definition of the GUI model, a common trait
among them is that the models are represented with
finite state machines with GUI layouts as states and
events as transitions.  There are two general
approaches of constructing the GUI model: static
approach or dynamic approach.  

An example of the static approach in constructing
the GUI model is
A\textsuperscript{3}E~\cite{azim2013_a3e}, 
which collects GUI states by
statically analyzing the program code, and uses taint
analysis to infer events that can cause transitions.
Other aforementioned tools collect GUI states by
extracting runtime GUI information during the execution
of the app, and capture GUI state transitions by
comparing the GUI states before and after exercising
each event.  A common limitation of these tools is that
event sequences are generated solely based on models
that only reflect GUI states transitions while ignoring
the internal states of the program.  As a result, in
their effort to avoid explosion, event sequences that
do not lead to new GUI states but trigger different
program behaviors may be overlooked.


In order to generate event sequences that are more
effective at exploring the program internal states,
several testing approaches such as
\textsc{collider}~\cite{collider},
\textsc{ACTeve}~\cite{anand2012_acteve}, etc., employ
symbolic execution to generate event sequences that are
distinguished by the program behaviors they can
trigger.  \textsc{Collider} uses symbolic execution to
discover fine-grained data dependencies between events,
and construct event sequences that can execute specific
program paths; \textsc{ACTeve} uses symbolic execution
to check whether an event's impact on program state is
relevant, and extends event sequences with only
relevant events.  Despite specific benefits and
drawbacks of symbolic execution, both approaches lack
a systematic GUI exploration strategy that threatens
the feasibility and effectiveness of testing. More specifically,
\textsc{Collider} assumes an existing GUI model of
the test app containing complete GUI transitions and
event handler mapping. However, constructing the GUI
model is not a trivial task, especially since \textsc{Collider}
targets apps that have complex user interaction
patterns. \textsc{ACTeve} on the other hand does not
use GUI model to generate events. Instead, it uses
symbolic execution to calculate the numeric values
of click event coordinates,
which results in excessive amount of symbolic execution
and inability to generate event sequences with length
more than four.

In this paper, we propose \emph{Android Path Explorer}
({\sc APEx}), an input generation framework aiming to
provide a systematic exploration and event sequence
generation for Android applications. {\sc APEx} is able to
generate not only a set of event sequences with high
code coverage, but also event sequences that can
trigger the execution of user-specified target code.
The framework is based on concolic execution that is
used to: (1) guide a systematic exploration of the
program behaviors and build an application model; and
(2) discover data dependencies between event handlers
and use the application model to construct concrete
event sequences.  Our work makes the following
contributions:

\begin{itemize} 

\item We propose a constraint-aware GUI Model that
indicates path constraints involved in GUI state
transitions, in order to support a systematic program
state exploration 

\item We propose a guided exploration algorithm that
exercises events in a prioritized order with the help
of concolic execution, to perform a systematic program
exploration and effective event sequence generation. 

\item We perform empirical evaluations to assess the
effectiveness of {\sc APEx} in generating event
sequences that can provide higher code coverage and
exercise specific execution paths. We also identify its
limitations.

\end{itemize}

The remainder of the paper is organized as follows.
In Section 2 we first explain relevant background
information and challenges in Android input generation.
In Section 3, we introduce our proposed constraint-aware GUI
model with formal definition.
We explain the details on {\sc APEx}'s
guided exploration algorithm and event sequence
generation process in Section 4. In Section 5 we evaluate the
performance of APEX and compare the performance of {\sc APEx}
with other available existing tools, and discuss the
limitations and future work of APEX. In Section 6 we
discuss related works in the area of Android testing
and input generation. We concludes the paper in Section 7.

\section{Challenges and Motivations}
\label{background}

The event-driven nature of Android application
framework introduces several challenges related to
input generation. In this section, we first discuss
these challenges, provide an overview of existing
approaches of using GUI models to overcome these
challenges, and finally, introduce a constraint-aware
GUI model structure that we use in {\sc APEx} to address the
challenges and limitations of existing tools.

\subsection{Android Background and Challenges}
\label{section_2_challenge}

Android GUI applications are event-driven in nature.
Each application runs in an isolated process that has
its own VM. When an application is started, its main
thread runs in a loop that listens for events, and
triggers corresponding callback methods defined in the
program code or in the Android framework.  Typically,
an event can be captured from a direct user input such
as tapping on a button or typing into a text field.
Events can also be generated from the system, including
a system wide broadcast of battery status change.
Callback methods for corresponding events are called
event handlers.

The Android application framework provides several
application components that serve as different ways of
interacting with the system.  \emph{Activities} is the
type of component that provides user interfaces on the
device screen.  A typical GUI application usually
contains several activities and one main activity which
serves as the main entry point when the user starts the
application.  Each activity is declared in the
application manifest file, \textit{AndroidManifest},
which indicates the main activity and specifies the
ways of interacting with each activity.  The user
interface on an activity is called a layout, which can
be either statically declared in an XML file, or
defined in the program and inflated during runtime.  An
activity class defines a series of the activity's life
cycle callbacks and event handlers for the GUI
components within its layout.



For systematic input generation
approaches,
we identify the following three main challenges:
\begin{itemize}

\item 
\textbf{Extracting UI information such as event
parameters, event handler registrations.} Due to the
design freedom granted to Android app developers,
extracting events and identifying event handlers is not
a trivial task. For instance, there are two options
to register an event handler in Android.
Assume a Button widget named \textsf{b1} within layout
\textsf{L1}, \textsf{b1}'s \textsf{onClick} event
handler can be registered in: (1) \textsf{L1}'s XML
declaration file (if \textsf{L1} is declared in XML)
using \textsf{android:onClick=``onClick1''}, where the
name \textsf{onClick1} implies a method with signature
of \textsf{``void onClick1(View v)''} is defined within
the activity class that will load the layout
\textsf{L1}, or, (2) somewhere in the activity class
using \textsf{b1.setOnClickListener()}. Due to the
possibility that a layout can be loaded into different
activities, in the first case, there will be multiple
\textsf{onClick1} method instances in each of the
activity classes. Moreover, \textsf{b1} can change its
\textsf{onClick} event handler during runtime using the
second way of registration. Although the above example
is not commonly seen in regular apps, an effective
input generation tool should have the ability to
correctly extract such UI information.

\item
\textbf{Handling implicit callbacks.} Implicit
callbacks exist in the control flow of Android
Application Framework which is outside the scope of the
program code. For example, when a ``click'' event is
applied to button \textsf{b1}, we can anticipate the
execution of event handler \textsf{onClick}.  However,
if the click event changes the device screen from
activity \textsf{A1} to \textsf{A2}, several life cycle
callbacks of \textsf{A1} and \textsf{A2} will also be
executed, which results in the execution of
\{\textsf{onClick, A1.onPause, A1.onStop,
A2.onCreate}\}. Such control flow is implicit and not
retrievable by simple static analysis on the program
code. Yet it is important for an input generation tool
to identify the implicit callbacks, in order to gain a
better understanding of the application behaviors.

\item
\textbf{Avoiding explosion of the number of event
sequences.} In theory, there exists a finite number of
event sequences that can trigger the execution of all
the reachable code in a program. However, it is
impractical to precisely generate such set of event
sequences in a larger-scale application in a timely
manner, due to the computational complexity of required
code analyses.  The more practical approach is to
produce an over-approximation of the necessary event
sequences that can be generated in a reasonable amount
of time. However, the number of feasible event
sequences can grow exponentially with the number of
available events. Therefore, an input generation tool
must maintain a balance between code coverage and the
number of event sequences.

\end{itemize}

Next, we discuss the effectiveness and limitations of
using GUI models to address the above-mentioned
challenges.

\subsection{Usage of GUI Models in Input Generation}



Most of existing model-based~\cite{amalfitano2015_mobiguitar}~\cite{azim2013_a3e}~\cite{choi2013_swifthand}~\cite{yang2013_orbit}
and systematic~\cite{mahmood2014_evodroid}~\cite{collider}
input generation tools utilize GUI models to
generate event sequences.  A GUI model is the
abstraction of an application's user interface,
including activities, layouts, and transition relations
between activities.  A commonly used GUI model
representation is \emph{GUI Transition Graph}~\cite{azim2013_a3e}, e.g.,
\(G=(V,E)\), where the vertices \(V\) are abstractions of
the activity states, usually represented by
a hierarchy of layouts and widgets
and their corresponding events, while the edges \(E\) represent
the events that trigger transitions between activities.


Model-based input generation tools build the GUI model
either by static analysis or by dynamic exploration
(e.g., depth-first exploration), and generate event
sequences along the GUI model building process.  Some
systematic input generation tools analyze the event
handlers in the GUI model with more complex analysis
techniques such as symbolic execution~\cite{king1976_symbolic} and
taint analysis~\cite{azim2013_a3e} to generate additional event
sequences that expose specific program behaviors.

While the model-based input generation tools can
effectively traverse the application's GUI states, they
are unreliable in exposing program behaviors underneath
the user interface.  On the other hand, while
systematic input generation tools are capable of
generating event sequences that trigger specific
program behaviors, their event sequence generation is
still based on the same GUI model structure which only
reflect GUI state transitions. The separation of GUI
model construction and event sequence generation
decreases the overall efficiency of the existing
systematic input generation tools.


\begin{figure}[t]
	\begin{subfigure}[b]{0.3\textwidth}
		\includegraphics[width=\textwidth]{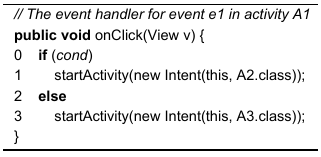}
		\caption{Event handler method in Java code}
		\label{fig:impreciseGUI:code}
	\end{subfigure}
	\begin{subfigure}[b]{0.16\textwidth}
		\includegraphics[width=\textwidth]{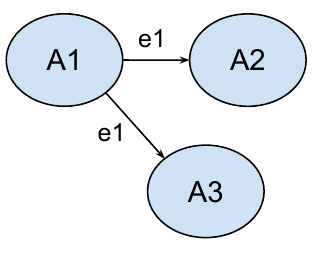}
		\caption{GUI Model}
		\label{fig:impreciseGUI:model}
	\end{subfigure}
	\begin{subfigure}[b]{0.5\textwidth}
		\includegraphics[width=\textwidth]{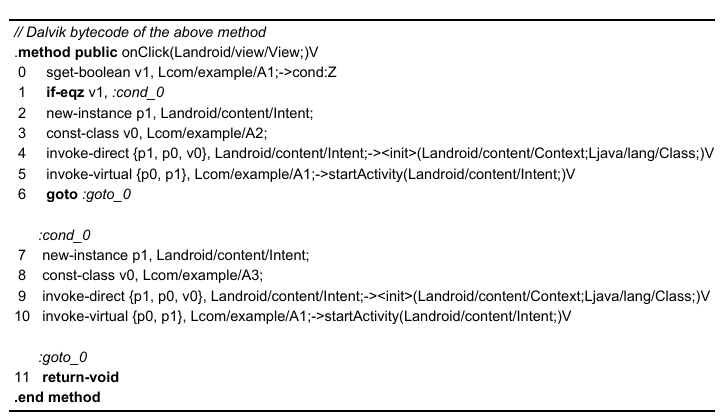}
		\caption{Event handler method in Dalvik bytecode}
		\label{fig:impreciseGUI:bytecode}
	\end{subfigure}
	\caption{An example of imprecise GUI state transitions in GUI Models}
	\label{fig:impreciseGUI}
\end{figure}

Furthermore, a common limitation of the above mentioned
tools is the imprecise state transition representations
in their GUI models.  Figure~\ref{fig:impreciseGUI}
shows a simple example of when this imprecision can
occur. As we can see in the event handler method in
Figure~\ref{fig:impreciseGUI:code}, there are two
possible activity transition outcomes by applying event
\textsf{e1} depending on the path condition
\textit{cond}. Obviously, the event sequence to reach
$A2$ and the event sequence to reach $A3$ would be
different. However, the two outgoing edges in the
resulting GUI model in
Figure~\ref{fig:impreciseGUI:model} are the same : $A1
\xrightarrow{e1} A2$ and $A1 \xrightarrow{e1} A3$.  In
order to find the correct event sequence for each GUI
transition, extra analysis effort will be required,
which undermines the efficiency and effectiveness of
event sequence generation.

\section{Constraint-Aware GUI Model}
\label{model_definition}

To address the limitation of imprecise state transitions, we propose an enhanced GUI model structure, named
\textit{constraint-aware GUI Model}, that uses a
more fine-grained representation for the GUI state transitions.
The main
difference between Constraint-Aware GUI Model and the
traditional GUI model is that, Constraint-Aware GUI
Model uses \textit{event summaries} to represent GUI
state transitions instead of using \textit{events}.
Formally, a constraint-aware GUI Model $M$ is a tuple:
\[ M = (S, S^*, E, H, P, \theta, \lambda, \Sigma , T) \] where
\begin{itemize} {

\item{\(S\) is a set of GUI states}

\item{\(S^* \subseteq S \) are the initial GUI states}

\item{\(E\) is a set of events} 

\item{\(H\) is a set of event handler methods}

\item{\(P\) is a set of program execution paths where
each path is a sequential list of bytecode statements.}

\item{\(\theta : E \rightarrow H\) is an event
handler mapping function.  $\theta(e)$ represents a set
of event handlers that will be executed after applying
event $e$}

\item{\(\lambda : H \rightarrow P\) is a path mapping
function. $\lambda(h)$ represents the paths in
the inter-procedural control flow graph that
start with the first statement of $h$ and end with
the return statement of $h$}

\item{\(\Sigma = E \times P\) is a set of event
summaries}

\item{\(T \subseteq S \times \Sigma \times S\) is the
transition relation between one GUI state to another}

}

\end{itemize}

Each GUI state \(s \in S\) represents a unique GUI
layout containing a set of events. 
An \textit{event} \(e \in E\)
can be either a \textit{user event} or a \textit{system event}.
In the scope of this paper, user events include:
tapping, swiping, long clicking, text, and other
special key events~\cite{keyevent} such as \textit{Home},
\textit{Back}, etc.; and system events include all
the system broadcasts (e.g. \textit{AIRPLANE\_MODE},
\textit{HEADSET\_PLUG}, etc.) that can
be emulated through the Android activity manager
program~\cite{am} using Intent~\cite{intent} objects.

\subsection{Entry Events}
\label{s_entryevents}
\begin{figure}[t]
	\includegraphics[width=0.5\textwidth]{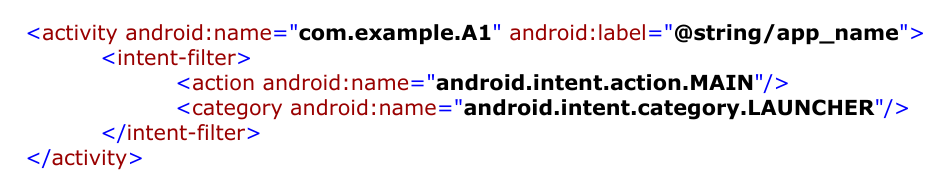}
	\caption{Declaration of ``main activity'' entry event}
	\label{fig:entry:main}
\end{figure}

\begin{figure}[t]
	\includegraphics[width=0.5\textwidth]{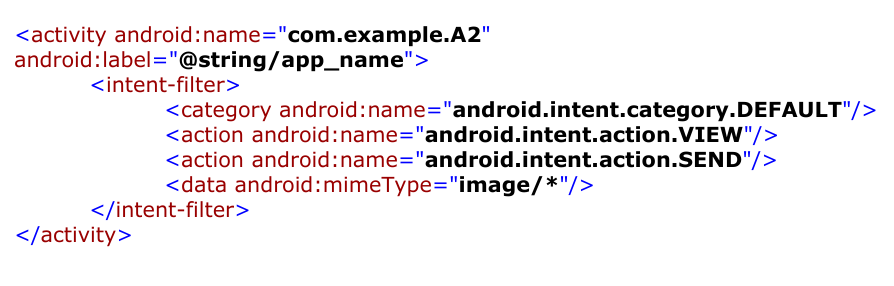}
	\caption{Receiver for a system broadcast that can start an activity with ``View'' or ``Send'' action}
	\label{fig:entry:action}
\end{figure}

Entry events are the system events that can start an app's functions from external sources outside of the app. There are two types of \textit{entry events}: (1) events that trigger specific activities or services of a specific app, (2) events that send broadcasts to all the apps that registered corresponding broadcast receivers. The main difference between the two types is, type 1 entry event has a specific target activity or service, while type 2 entry event might be received by multiple app components in different apps.
Both types of entry events can be found in the \textit{AndroidManifest.xml} files, declared with tag name $\langle intent-filter \rangle$ within the $\langle activity \rangle$, $\langle service \rangle$, and $\langle receiver \rangle$ nodes. As an example of type 1 entry event, every Android app that has a graphical user interface must label an activity as the ``Main Activity'', as shown in Figure~\ref{fig:entry:main}. The value ``android.intent.action.MAIN'' indicates that this activity can be started with command ``\textit{am start -n com.example/com.example.A1}''. Activities other than the main activity may also be started externally, using type (2) entry events. Figure~\ref{fig:entry:action} shows such an activity that can be started by a broadcast intent with action ``View'' or ``Send'' along with an image file. Note that not all activities can be started externally, only those with properly defined intent filters can be started by the corresponding external intents.

\begin{figure*}[t]
	\begin{subfigure}[b]{0.4\textwidth}
		\includegraphics[width=\textwidth]{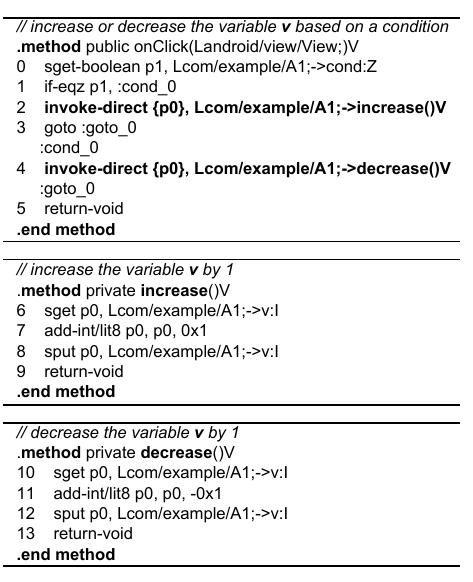}
		\caption{Definition of 3 methods, \textbf{onClick}, \textbf{increase}, and \textbf{decrease}.}
		\label{fig:ipcfg:methods}
	\end{subfigure}
	\begin{subfigure}[b]{0.5\textwidth}
		\includegraphics[width=\textwidth]{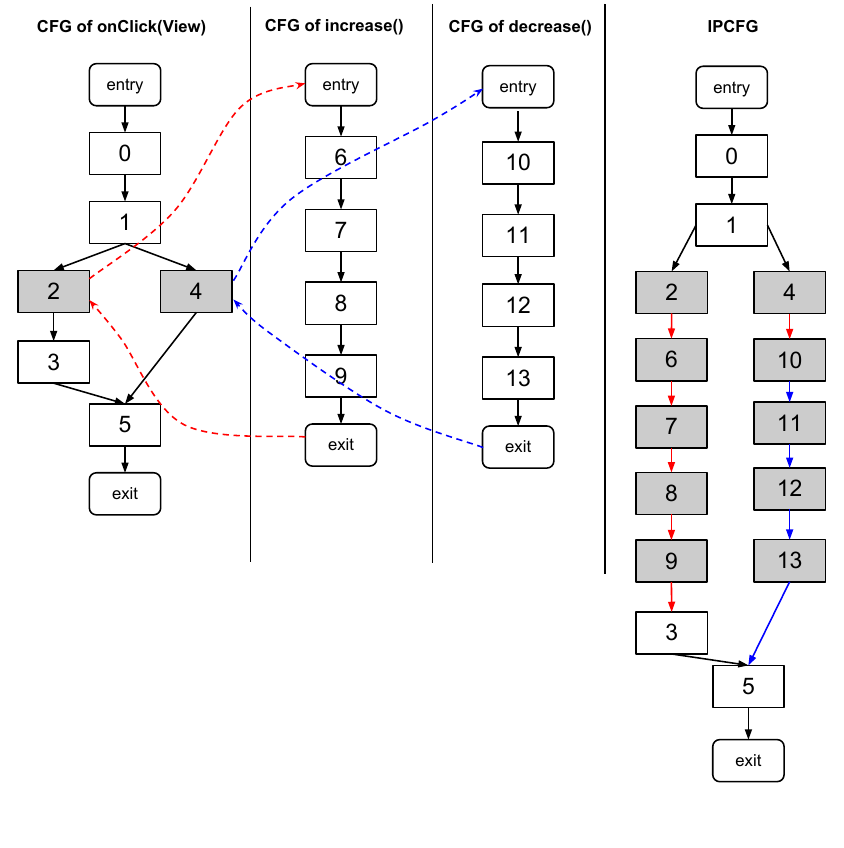}
		\caption{The CFG for each method and the IPCFG for \textbf{onClick}}
		\label{fig:ipcfg:graphs}
	\end{subfigure}
	\caption{An example of Inter-procedural Control Flow Graph}
	\label{fig:ipcfg}
\end{figure*}

\subsection{Source GUI State}
With the exception of entry events, every available event
must have a \textit{source GUI state}.
A user event's source GUI state is the layout in which
the corresponding GUI component is defined.
A non-entry system
event's source GUI state is the state when the
\textit{BroadcastReceiver}~\cite{broadcastreceiver}
corresponding to the event is registered.
This is because the broadcast receivers of 
non-entry system events are not statically registered in
the AndroidManifest, but dynamically registered in
the program code using API such as ``Context.registerReceiver(...)''. Since such system events
can only work after the receiver registration code is executed,
we consider the GUI state at the time of execution as their source GUI state.

\subsection{Event Summary}
An \textit{event summary} \(\sigma \in \Sigma\) is a 2-tuple containing
an event and a specific program execution path in its
event handler method. The transition between two
states is represented by an event summary. For the
example in Figure~\ref{fig:impreciseGUI}, a
constraint-aware GUI model will have two state
transitions: $A1 \xrightarrow{\sigma1} A2$ and $A1
\xrightarrow{\sigma2} A3$, where $\sigma1 = (e1,
A1.onClick:\{0,1,2,3,4,5,6,11\})$ and $\sigma2 = (e1,
A1.onClick:\{0,1,7,8,9,10,11\})$ (the line indices refer to the bytecode in Figure~\ref{fig:impreciseGUI:bytecode}). An event
summary is \textit{concrete} if its program path has been concretely
executed; it is \textit{symbolic} if its program path has not yet been executed. Symbolic event summaries are generated for the unexplored paths in the inter-procedural control flow graph. The details in generation and solving of symbolic event summaries are explained in Section~\ref{section_3_generate_ses}.

The advantages of representing GUI state transition
with event summaries instead of events are:
\begin{enumerate}

\item the application's user interface is better
represented since the state transitions are more
precise using event summaries,

\item the GUI exploration process can easily identify
unexplored GUI state transitions from the unexecuted
paths in the event handler methods, and explore towards
these GUI states,

\item the event sequence generation can narrow its
search space and remove infeasible event sequences by
inferring data dependencies and control flows from the
event summaries.

\end{enumerate}

\subsection{Inter-Procedural Control Flow Graph}
The \textit{inter-procedural control flow graph} (IPCFG)~\cite{ipcfg} is a graph that combines the control flow graph of each method by connecting the method entries and exits with their callsites. 
Figure~\ref{fig:ipcfg} shows an example of the IPCFG. Figure~\ref{fig:ipcfg:methods} shows the dalvik bytecode for three methods: \textit{onClick(View)}, \textit{increase()}, and \textit{decrease()}. As we can see in Figure~\ref{fig:ipcfg:graphs}, there are two paths in the control flow graph of method onClick. One path contains statement 2 that invokes method \textit{increase()}, while the other contains statement 4 that invokes method \textit{decrease()}. To build the IPCFG, we connect the entries and exits of the two invoked methods to their callsites, resulting in an IPCFG that contains all the statements. In the context of this paper, the IPCFG is consist of event handler methods, system event callbacks, and all the invoked methods within them, recursively.

\section{Introducing APEX}
\label{approach}

The two main goals of {\sc APEX} are:(1) to provide a
systematic exploration of an Android application, and
(2) to generate event sequences that can expose
different program behaviors and trigger the execution
of user specified code targets.
To achieve these goals, {\sc APEX} uses a guided GUI exploration strategy to perform systematic exploration
of the program behaviors and build a constraint-aware GUI
model. The GUI model is then used to discover data dependencies between event handlers and generate new event
sequences that expose more program behaviors and GUI transitions. Symbolic execution is used to infer path
constraints and construct event sequences.

Next we explain the details of the guided GUI
exploration strategy and the event sequence generation.

\subsection{Components of the GUI Exploration}

Our guided GUI exploration algorithm combines dynamic
GUI exploration with concolic execution to generate
event sequences that can lead the exploration to new
program behaviors. Algorithm~\ref{alg_exploration}
shows an overview
of the proposed guided GUI exploration algorithm. 

\renewcommand{\algorithmicforall}{\textbf{for each}}

\begin{algorithm*}[t]
\caption{Guided GUI Exploration Algorithm}
\label{alg_exploration}
\hspace*{\algorithmicindent} \textbf{Input}: Test app \textit{A}, code targets \textit{T}\\
\hspace*{\algorithmicindent} \textbf{Output}: GUI model \textit{M}, exploration history \textit{h} 
\begin{algorithmic}[1]
	\State $M \gets \phi$	\Comment{$M$ is the GUI model}
	\State $Q \gets \Call{GetEntryEvents}{A}$	\Comment{$Q$ is the event sequence priority queue}
	\State $L \gets \phi$	\Comment{$L$ is the symbolic event summary priority queue}
	\State $history \gets \phi$
	\State $icfg \gets interproceduralCFG(A)$	\Comment{$icfg$ is the inter-procedural control flow graph}
	\State instrument and install A
	\While{$Q$ is not empty $or$ $L$ is not empty} 
	\Comment{Termination condition of the exploration process}
		\While{$Q$ is not empty}	\Comment execute all the event sequences in the queue
			\State $seq \gets \Call{DeQueue}{Q}$
			\State $(layout, handlers, exec\_log) \gets \Call{Apply}{seq}$
						\Comment{Execute an event sequence and extract runtime information}
			\State add$(seq, layout, handlers, log)$ to $history$
			\State $\sigma \gets (\Call{FinalEvent}{seq}, exec\_log)$ \Comment{Create an event summary}
			\State add $\sigma$ to $M$
			\If{$layout \not\in M$}		
				\State $events \gets \Call{extractEvents}{layout}$	\Comment{Extract events from new GUI state}
				\State $Q \gets \Call{update}{Q, events, T}$		\Comment Add extracted events to queue
			\EndIf
			\State $M \gets \Call{update}{M, \sigma, layout}$	\Comment Update the GUI model(see Section~\ref{section_update_model})
			\State $sp \gets \Call{GetSymbolicPaths}{\sigma, icfg}$
			\State $L \gets \Call{update}{L, sp, T}$		\Comment Add symbolic event summaries to queue(see Section~\ref{section_3_generate_ses})
		\EndWhile
		
		\State
		\If{$L$ is not empty}
			\State $summary \gets \Call{DeQueue}{L}$
			\State $seq \gets \Call{SequenceGen}{M, L, summary}$ 
				\Comment{Event sequences generation (see Section~\ref{section_seqgen})}
			\State add $seq$ to $Q$
		\EndIf
	\EndWhile
	\State \Call{Return}{$M, h$}

\end{algorithmic}
\end{algorithm*}

%
%
%
%

The algorithm starts with a test app and 
an optional input that is the
user-specified code targets, and ends when no new event
sequences can be generated and explored.  
User-specified code targets can be either
line numbers in the source code or the bytecode indices
of dalvik bytecode statements. Although APEX only works
on the dalvik bytecode, it can still take source
code line number as input. This is because the Android
program binaries by default keeps the source line
number as part of the debug information, which can be
easily extracted by reverse engineering tools such as
apktool~\cite{apktool}. Using the extracted debug information,
we can then convert the source code line numbers to corresponding bytecode indices. The code targets
play an important role in the prioritization mechanism that
leads the GUI exploration towards execution of
these code targets.

The exploration process maintains three data structures: a
constraint-aware GUI model \textit{M}, 
an event sequence queue \textit{Q},
and a symbolic event summary queue \textit{L}.

\noindent
\textbf{Constraint-Aware GUI model} has been introduced
in Section~\ref{model_definition}. We use this model to
provide an abstraction of the application's user
interface with GUI states and GUI state transitions
triggered by event summaries. The model is constructed
incrementally along the exploration, and is used by the
event sequence generator to construct event sequences.

\noindent \textbf{Event Sequence Queue} is a priority
queue that stores event sequence candidates
during the exploration. Each event sequence in the
queue is assigned a priority that represents its
potential in revealing new program behaviors and
triggering the execution of user-specified targets.
New event sequences are added to the queue in two
scenarios: (1) when the exploration arrives at an
unexplored GUI state, the available events are then
extracted from the layout hierarchy, and added to the
queue as partial event sequences; (2) when the event
sequence queue is empty, i.e., all the partial event
sequences in previously explored GUI states have been
exercised. When the individual events from existing
GUI states have all been executed, we then try to
generate new combinations of events using their
corresponding event summaries.

\noindent \textbf{Symbolic Event Summary Queue} is a
priority queue that stores the symbolic event
summaries during the exploration. A \textit{symbolic}
event summary is an event summary with an execution
path which has not been concretely executed. Whenever
an event sequence is applied during the exploration,
only one program path is executed in the event handlers
of the final event. We use inter-procedural 
Control Flow Graph to
extract the rest of the program paths, and create
symbolic event summaries for these paths. These
symbolic event summaries have the same event but
different path constraints, and therefore are
considered ``unsolved''. The unsolved symbolic event
summaries are added into this queue with a priority
that is determined by similar metrics of the event
sequence queue.

\subsection{Building Constraint-aware GUI Model}
\label{section_update_model}

Building the GUI model is an incremental process
throughout the course of exploration. As shown in
Algorithm~\ref{alg_exploration}, we begin the
exploration by initializing an empty GUI model (at line
1). The event sequence queue is initialized with
a set of entry points by analyzing the \textit{AndroidManifest}
file. In addition to the main activity of an app,
we also consider other activities that can be started
by a system broadcast event.
When multiple entry points are detected, multiple
starting events will be generated based on their specific
type of intent filters.
During the exploration, APEX executes the event sequences
based on their priority, and
extracts runtime information (line 10) including the
GUI layout, the event handler method, and the executed log.
The runtime information is used for GUI model updating,
event handler mapping, and event summary generation.

New GUI states are added in to the model when a new GUI
layout is discovered, i.e., the layout is not
\textit{equivalent} to any of the existing GUI states
in the model. We consider a layout \textit{l1} to be
equivalent to layout \textit{l2} when:

\begin{itemize}

	\item for each event $e \in l1$, there exists an event $e' \in l2$,
	such that: $\theta(e) = \theta(e')$,
	\item and vice versa.

\end{itemize}

When a new layout is discovered, a transition with the
event summary is created and added into the model (lines
12-13). Then, the events and their corresponding event handlers
are extracted to add into the event sequence queue $Q$ (lines 14-16).
We only create event summaries for the final
event of an event sequence to avoid redundancies. Using
the final event and the retrieved execution log, we can
build a \textit{concrete} event summary, indicating the
event sequence for this particular event summary has
been applied during the exploration, and therefore can
be easily recreated from the model. In the case where
there are unexplored branches in the event
handlers, we create symbolic event summaries for the
unexplored paths and add
them into a symbolic event summary queue (line 19-20),
which will be processed in the next phase.

When the exploration finishes execution of all
the event sequences in $Q$, it means the we have
traversed all the events from the known GUI states.
To continue the exploration towards unexplored
program behaviors and potentially undiscovered GUI
states, we move on to the next phase (lines 23-27)
where the symbolic event summary with highest priority
is processed to generate event sequence candidates.
Section
~\ref{section_seqgen} provides the details on how the
symbolic event summaries are processed.

After a new set of event sequences are added into
$Q$, the exploration phase (lines 8-21) can be resumed.
The exploration completes when both $Q$ and $L$ are empty,
meaning that all of the GUI states and program states
have been visited. Next, we explain the technical
details of the key components of APEX.

\subsection{Generating Event Parameters}
The first step of the exploration is to generate entry
events. We identify entry events from the AndroidManifest.xml
files, looking for \textit{activities}, \textit{services}, and \textit{receivers}
with statically declared intent filters. As explained in Section~\ref{s_entryevents}, most apps have at least one entry event that is the intent
for starting an app's main activity. For other statically registered entry events, we generate the corresponding parameters based on the intent action, category, and other relevant properties such as the data type, as shown in Figure~\ref{fig:entry:action}. These entry events are the initial members in the event sequence queue $Q$, and will executed using the activity manager (am)~\cite{am} program. 

During the exploration, non-entry system events may become available
via dynamically registered broadcast receivers. APEX identifies newly
available system events by monitoring the runtime execution log and searching
for method invocation statements calling the ``Context.registerReceiver(...)'' APIs. To generate parameters for these events, we use instrumentation to insert logging statements next to the receiver registration statements, to print out the parameters of the corresponding intent-filter when it's being registered. With that, we can generate event parameters accordingly.

For user events, we analyze the runtime layout hierarchy
using UIAutomator~\cite{uiautomator}, and collect all the
leaf nodes in the layout XML dump. We generate events
according to the widget types and screen coordinates from
the layout file. For text input widgets, we fill them with
randomly generated strings during first encounter. If the value
or format of the input text fails to satisfy certain path
conditions, the relevant path conditions will be recorded
in symbolic event summaries, which can be used to generate
new strings that satisfy the conditions. However, existing constraint
solvers have limited capability against constraints that involve
system API return values. To deal with this challenge, we
modeled several methods in the \textit{String} and \textit{StringBuilder} classes, including:
\textit{StringBuilder.append()}, \textit{String.equals()}, \textit{String.length()}, etc., to reduce the number of unsolvable path constraints.

Each of the available events in the newly discovered
layout will be put in a \textit{partial event sequence}
and then added to the event sequence queue.
A \textit{partial event sequence} contains only
one event. It is used as an extension of the current
event sequence which starts from the main entry and
ends at the source layout of the single event. Before
an event sequence is applied, the exploration first
check whether the source layout of the first event is
equivalent to current GUI state, to decide whether the
event sequence partial or full, and then proceed
accordingly. 

When the event sequence queue is empty,
it means the events in previously visited layouts have all
been applied at least once.  However, the GUI model at
this point might not be sound. The exploration could
have missed some GUI transitions that require a
specific event sequence to trigger.
To continue the exploration, the algorithm calls the sequence
generator to generate more event sequences using the
symbolic event summary queue. The details of the symbolic
execution component is in Section~\ref{section_seqgen}.
Through
the prioritization mechanism, the symbolic event
summaries whose execution path contains GUI transition
inducing statements will have higher priorities to be
solved. With the generated event sequences, the
exploration will be guided towards those hard-to-reach
GUI states. As a result, our exploration can construct
a solid constraint-aware GUI model which in turn
supports the event sequence generation process to yield
better results.

\subsection{Event Handler Mapping}
We use instrumentation to
map events to their corresponding event handlers during runtime.
Before testing, we instrument the app by
inserting logging statements at the beginning
and returning of each method, which prints out the method signature and a tag indicating the beginning or returning of this method.
During testing, we track the
method signature logs after applying each
event, and identify the event handler method as the 
root method of the execution log stack.
For the code examples in Figure~\ref{fig:ipcfg:methods}, the method signature log of \textit{onClick} could be: {\textit{onClick\_start}, \textit{increase\_start}, \textit{increase\_return}, \textit{onClick\_return}}. Based on the order of the log output, we can safely determine that this onClick method is the event handler
for the last executed event.


\subsection{Prioritization of Event Sequences}

{\sc APEX} uses event sequence prioritization to manage the
order in which the event sequences are applied. The
goal of our priority mechanism is to guide the
exploration process towards: (1) new GUI state
transitions, (2) early execution of user-specified
targets, and (3) unexecuted program paths. The
exploration process maintains an event sequence
priority queue to achieve this goal.  Two types of
event sequences are added into the priority queue
throughout the exploration. First, when a new GUI
layout is discovered, the available events within the
layout are extracted from the layout hierarchy, and
added into the queue as \textit{partial event
sequences}. Second, when the event sequence generator
solves a symbolic event summary, the resulting event
sequences are added into the queue as \textit{complete
event sequences}.

The event sequence queue re-prioritizes the event
sequences whenever new ones are added. We define the
following rules to help decide which event sequence has
the highest priority:

\begin{itemize}

\item Partial event sequences precedes complete event
sequences. This rule goes into effect in the scenario
where the event sequences provided by the sequence
generator result in a new GUI layout, which is the
goal of these complete event sequences. Naturally, the
newly discovered partial event sequences should be
prioritized in order to resume the exploration of the
program.

\item If two event sequences are both partial or both
complete, the one whose execution path contains more
targets has higher priority; if they contain same
amount of targets, the one whose execution path
contains GUI state transition related code has higher
priority.

\item If priorities between two event sequences are
still undecided, choose one arbitrarily to precede the
other.  

\end{itemize}

By prioritizing event sequences during the exploration,
we can effectively guide the exploration towards
unexposed program behaviors, which in turn helps to
avoid explosion in the number of event sequences.

\subsection{Event Sequence Generation}
\label{section_seqgen}
The event sequence generator is called when
all the
discovered events have been executed at least once,
i.e. when the event sequence queue is empty.
The goal of our event sequence generator is to
generate event sequences for the symbolic event
summaries generated along the exploration, in order
to guide the exploration towards new program behaviors.
As shown in
Algorithm~\ref{alg_exploration} line 16, when an event
sequence is executed, a set of symbolic event summaries
are generated along with the concrete event summary.
The sequence generation is accomplished using symbolic
execution along with the constraint-aware GUI
transition information provided by the GUI model. The
main challenge of using symbolic execution is the path
explosion problem~\cite{cadar2013symbolic}. We address this
challenge by prioritizing execution paths such that the
\emph{important} paths are symbolically executed first,
avoiding unnecessary symbolic executions.

\subsubsection{Generating Symbolic Event Summaries}
\label{section_3_generate_ses}
Symbolic event summaries are generated for
event handlers that have multiple execution paths,
using inter-procedural control
flow graphs~\cite{bodik1997interprocedural}. 
An IPCFG
is a graph that combines the CFGs of all methods by
connecting method entries and exits with their call sites.
We first construct the IPCFG only for the event handler,
then collect all the paths within the graph,
and identify the concretely executed path
based on runtime execution logs. For each non-executed
path, we pair it with the corresponding event and
create a symbolic event summary.
To deal with the potential path explosion problem,
we use a prioritization mechanism to ensure the
event summaries with more ``relevant" paths are processed
first. We provide more details on the prioritization in
Section~\ref{section_3_symbolic_prioritization}

\subsubsection{Symbolic Execution}


In symbolic execution, \emph{symbolic states}
are the states or values of global variables represented
by symbols. Since event handler methods in Android generally
have one single parameter, which is a \emph{View}
object correlating to the event, only global variables
(usually field members of global objects) need to be
symbolized.The \emph{path constraints} are a set of
constraints that must all be satisfied to enable
the execution of a program path.
We use symbolic execution to find event sequences
for previously un-covered execution paths.

\renewcommand{\algorithmicforall}{\textbf{for each}}
\algnewcommand\And{\textbf{and}}

\begin{algorithm}[t]
\caption{Symbolic Execution Algorithm}
\label{alg_sym}
\hspace*{\algorithmicindent} \textbf{Input}: Initial symbolic states $S_0$, execution path $p$\\
\hspace*{\algorithmicindent} \textbf{Output}: symbolic states \textit{S}, path condition \textit{C} 
\begin{algorithmic}[1]
	\State $S \gets S_0$	\Comment initialize symbolic states
	\State $C \gets true$	\Comment initialize path condition
	\ForAll{statement $s$ in $p$}
		\If{$s$ is first in a block}	\Comment new path condition
			\State $\sigma \gets \Call{GenerateConstraint}{S, s}$
			\State $C \gets C\And\sigma$
		\EndIf
		\State interpret $s$ and update $S$	\Comment update symbolic states
	\EndFor

\end{algorithmic}
\end{algorithm}

Algorithm~\ref{alg_sym} shows the basic workflow
of our symbolic execution process.
The algorithm takes a set of initial symbolic states and
a program path \textit{p} as input,
uses symbolic values to represent
method input parameters and global variables, and
execute each statement in \textit{p} sequentially
to update the symbolic states and path conditions,
until reaching the end of the execution path. The
algorithm returns the updated
symbolic states and path constraints.

\begin{figure}[t]
	\includegraphics[width=0.5\textwidth]{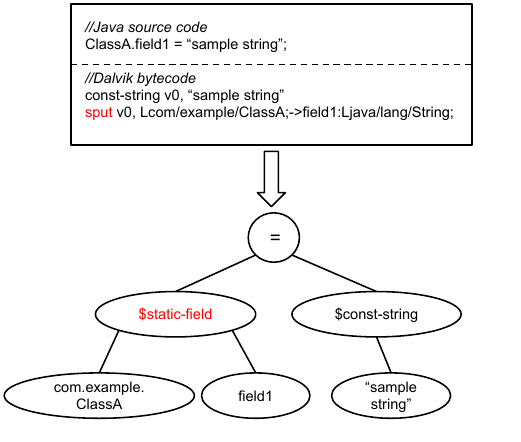}
	\caption{An visualization of parsing Dalvik bytecode into AST format}
	\label{fig:apex_ExpressionExample}
\end{figure}

In our implementation, the \emph{symbolic states} and
\emph{path constraints} are represented in the form of
Abstract Syntax Trees (AST), using keywords to
indicate symbols.  The Dalvik bytecode instruction
set~\cite{bytecode} contains 219 different
instructions. Among them are many instructions that
perform the same function but reflects different
operand sizes or data types. For example, there are 7
instructions for loading an element from an array:
\emph{aget, aget-wide, aget-object, aget-Boolean,
aget-byte, aget-char, aget-short}.  Our symbolic
execution parses these instructions using the same
keyword \emph{\$aget}. We have created 17 different
keywords for the whole Dalvik bytecode instruction set.
Figure~\ref{fig:apex_ExpressionExample} shows an
example of the symbolic state expression format of
bytecode instruction \emph{sput}, which writes a value
to a static field. In this example, the root node of
the AST has the name ``='', indicating this expression
is a symbolic state. The left child of the root node
has a keyword \emph{\$static-field}, representing a
symbolic value using \emph{com.example.ClassA.field1}
as its unique signature.

We have implemented the symbolic execution in a VM like
structure. The symbolic VM contains heap and method
stack. In the method stack, each method is assigned a
set of registers that are used to stores local
variables. The value stored in a register can be either
a \emph{literal value} or a \emph{reference value}.
Reference values usually represent the \emph{address}
of an object from the heap. We implemented the heap as
a list of objects with a symbolic value and field
members.  At the end of each symbolic execution, the
state of global variables are collected from the heap.

Due to the event-driven nature, solving a path constraint
in an event handler method often requires a series of
other event handler methods to be executed first, in order
to change the value of symbolic states to satisfy the path
constraint. Symbolic execution in Android programs is not
about ``tuning'' the input parameters of a function, but rather
a recursive process of looking into the path summaries of other
event handlers and construct event sequences that can satisfy
the path constraints.

Next, these event handlers are mapped back to events. With
the GUI model, we can eventually construct an
event sequence that starts from the application main
entry point and connects those necessary events and
ends with the target event.

\subsubsection{Constraint Solver}

Our constraint solving process takes the GUI model and
a symbolic event summary as input, outputs a list of
event sequences candidates that can potential turn
the symbolic event summary into a concrete one.
This is achieved by finding the existing event summaries
that can change
system states in certain ways so that the path constraint
of the target path can be satisfied. Algorithm~\ref{alg_solver}
shows the details of the constraint solving process.

\renewcommand{\algorithmicforall}{\textbf{for each}}

\begin{algorithm}[t]
\caption{Constraint Solving Algorithm}
\label{alg_solver}
\hspace*{\algorithmicindent} 
\textbf{Input}: GUI Model $M$, Symbolic event summary $\sigma$\\
\hspace*{\algorithmicindent} 
\textbf{Output}: Event sequence list $L$
\begin{algorithmic}[1]
	\State $L \gets \phi$
	\State $c \gets \Call{SymbolicExecution}{\sigma}$	\Comment get path constraints
	\State $e \gets \Call{Event}{\sigma}$
	\ForAll{event summary $s$ in $M$}
		\If{\Call{Satisfy}{s, c}}	
			\State $seq \gets \Call{FindPath}{M,s}$	\Comment get partial sequence
			\State append $e$ to $seq$				\Comment complete sequence
			\State add $seq$ to $L$
		\EndIf
	\EndFor
	\State return $L$
\end{algorithmic}
\end{algorithm}

Generally, the constraint solving process first searches for
relevant symbolic states from other path summaries that
can potentially satisfy each constraints, then the
SMT solver~\cite{cvc4} is used to
determine whether the relevant symbolic states satisfy
the path constraints.
When path summaries whose
symbolic states satisfy all the path constraints are
found, the event sequences of these path summaries is
inserted in the front of the existing event sequence.
These newly found path summaries then become the new
subject of constraint solving. This process repeats
until there are no new path constraints to solve.
As a result, a list of event sequences are generated.

In practice, the SMT solvers are not suitable for
checking certain types of path constraints. These path
constrains can be categorized into two groups:
\begin{enumerate}
	\item implicit constraints that relates to system events
	\item constraints involving the return value of APIs
\end{enumerate}

We deal with these path constraints by modeling the
most frequently encountered APIs, including methods of
\textit{Intent}, \textit{String}, etc.
For example, the Intent object used to start an activity
can contain extra data which would determine how the activity
reacts. We model the \textit{Intent.getExtra()}
and \textit{Intent.getExtraString()} so that when these methods
appear in a path constraint, we can skip the SMT solver
and directly generate a system broadcast event with
correct parameters.

To deal with path constraints involving String values,
e.g. string values from EditText widgets,
we model several methods of the \textit{StringBuilder}
and \textit{String} class so the symbolic execution 
component can generate string values usable in the 
\textit{cvc4}~\cite{cvc4} SMT solver.

\subsection{Prioritization of Symbolic Event Summaries}
\label{section_3_symbolic_prioritization}
In addition to dealing with the path explosion problem, the purpose of prioritizing symbolic event summaries in the
queue $L$ is to ensure that, the event summary that is most likely to enhance the GUI model gets processed first. As shown
in Algorithm~\ref{alg_exploration}, after the execution of an event,
APEX collect the runtime execution log to create a concrete event summary, and then collect all the unexplored program paths with the same method entry point to create symbolic event summaries. Symbolic event summaries are stored in
a priority queue, with different priority values, based on the following rules:
\begin{enumerate}
\item The event summary that contains more
user-specified targets has higher priority.
\item The event summary that contains GUI transition
related code has higher priority.
\item The event summary that contains more ``write
field'' operations: \emph{sput} and \emph{iput} has
higher priority.

\end{enumerate} 
Rule 1 adds priority to the event summaries that contain user specified code targets. This rule ensures the early execution of code targets, increasing the efficiency of APEX's targeted input generation. Rule 2 adds priority to event summaries that are likely to expose new GUI states. Rule 3 add priority to those that are more valuable in terms of event sequence generation. The two operator \textit{sput} and \textit{iput}, are field assigning operators. As an example, the statement \textit{sput v0, Lcom/example/A1;->x:I} means assigning the value of register \textit{v0} to the static field \textit{A1.x}, and\textit{iput} is used when assigning values to an object's instance fields. These two operators are valuable because the symbolic execution on their program paths will likely lead to changes of symbolic states, which can be helpful in providing partial solutions to other symbolic event summaries.

We implement the prioritization mechanism by assign a priority value to each symbolic event summary in the queue. Event summaries that have higher priorities according to the rules will have a higher number.
Furthermore, we include a penalty mechanism, to prevent unsolvable event summaries that have high priority from occupying the top spots. When a selected event summary cannot be solved, it will be ineligible for being selected for the next several iterations.

\section{Evaluation}

In this section, we evaluate the performance of APEX
in terms of code coverage and effectiveness of 
targeted input generation. We aim to answer the 
following research questions:
\begin{itemize}
	\item \textbf{RQ1: Code Coverage.} 
	How does APEX compare with existing 
	tools in terms of cover coverage?
	\item \textbf{RQ2: Targeted Input Generation.} 
	How effective is APEX in generating 
	input for specific code targets?
\end{itemize}

To answer RQ1, we test APEX on 48 real world apps, and
compare the code coverage with state-of-the-art tools:
\textit{Monkey}~\cite{monkey},
\textit{Sapienz}~\cite{sapienz}, and
\textit{Stoat}~\cite{stoat}.  To answer RQ2, we test
APEX on 8 benchmark apps consisting of five malware
samples, two benign apps that have been used by our
baseline system, Colider~\cite{collider}, and a
microbenchmark program, Dragon, that was designed to
contain many hard-to-reach path targets.  We also
compare the target coverage with concolic execution
based approach \textit{Collider} by using the two apps
that have been used in their evaluation.

\subsection{Experimental Setup} \textbf{Apps Under
Test.} For the code coverage evaluation (RQ1), we
collected 48 test apps, including 44 open-source apps
from the F-Droid repository~\cite{fdroid} and 4 Play
Store apps: \textit{Lolcat}, \textit{Random Music
Player}, \textit{Wikipedia}, and \textit{Wordpress}.
These test apps belong to various categories such as
entertainment, productivity, news, etc., and have been
tested in several previous works~\cite{arewethereyet,
sapienz, stoat}, including sapienz and stoat.  For the
targeted input generation evaluation (RQ2), we used
eight apps, including five malware samples from DARPA
APAC engagements, \textit{BattleStat},
\textit{rLurker}, \textit{AudioSideKick},
\textit{AWeather}, and \textit{Engologist}. We also
included two real world apps, \textit{TippyTipper} and
\textit{MunchLife}, and a microbenchmark
app, \textit{Dragon}, 

\textbf{Measuring Code Coverage.}
In this evaluation, we use the number of Dalvik 
bytecode lines to calculate the code coverage. 
Comparing to method coverage, this 
metric can better represent the percentage of 
different program paths being explored. 
The test apps are instrumented prior to testing 
to enable the measurement of code coverage.  
Logging statements were inserted at the beginning 
of each method, returning of each method, and the
beginning of each bytecode block. During testing, 
we use logcat to monitor the system log output 
to capture the blocks that were executed, 
and count the total number of bytecode 
instructions in those blocks to calculate an 
overall code coverage.

\textbf{Testing Methodology.}
To answer RQ1, we test APEX on 48 real world apps 
without any code targets as input, and limit its 
run time to maximum 60 minutes per application.
To compare the code coverage, we then test Monkey
on the same apps.
For monkey, we specified its \textit{ALLOWED-PACKAGE}
field to ensure that monkey focuses on the test app, 
and enabled the \textit{ignore-crashes} and 
\textit{ignore-security-exceptions} flags so that 
monkey can still report crashes and security 
exceptions without stopping prematurely. 
We set monkey's total event count to 100,000 for 
each app, as we observed that monkey reached its 
maximum code coverage on all the test
apps well before it finishes firing all the events.
%
%
%
%
To compare the code coverage with sapienz and 
stoat, we use their published results instead 
of running them, for the following reasons:
\begin{enumerate}

\item Sapienz and Stoat require the test apps to be
instrumented with Ella~\cite{ella} which is different
from APEX’s instrumentation component, and their input
generation processes benefit from runtime code coverage
that cannot be provided by APEX instrumented apps.

\item Based on our experience, code coverage is a fairly consistent metric, especially given a long testing time. Since both Sapienz and Stoat run each test app for 1 hour, we consider their published code coverage result to be consistent and fair to be compared with.

\end{enumerate}

\begin{table*}[t]
\centering
\caption{Overall Code Coverage Results}
\label{apex:table_cc}
	\begin{tabular}{c|l|lccc|llll}
\hline
\multirow{2}{*}{\textbf{\begin{tabular}[c]{@{}c@{}}\# of\\ Apps\end{tabular}}} & \multicolumn{5}{c|}{\textbf{App Under Test}} & \multicolumn{4}{c}{\textbf{Code Coverage (\%)}} \\ \cline{2-10} 
 & \textbf{App name} & \textbf{LOC} & \multicolumn{1}{l}{\textbf{\# activities}} & \multicolumn{1}{l}{\textbf{\# widgets}} & \multicolumn{1}{l|}{\textbf{\# event handlers}} & \textbf{M} & \textbf{SA} & \textbf{ST} & \textbf{AP} \\ \hline
\multirow{18}{*}{\textbf{18}} & ADSdroid & 35054 & 2 & 6 & 9 & 60 & 38 & 28 & \textbf{84} \\ \cline{2-10} 
 & Lolcat & 2942 & 1 & 10 & 23 & 23 & 18 & 24 & \textbf{63} \\ \cline{2-10} 
 & Mirrored & 3836 & 4 & 8 & 39 & 69 & 33 & 50 & \textbf{77} \\ \cline{2-10} 
 & Hot Death & 19336 & 3 & 11 & 33 & 60 & 57 & 70 & \textbf{95} \\ \cline{2-10} 
 & HNdroid & 17058 & 5 & 19 & 47 & 22 & 11 & 10 & \textbf{31} \\ \cline{2-10} 
 & swiftp & 12339 & 3 & 134 & 51 & 24 & 13 & 18 & \textbf{32} \\ \cline{2-10} 
 & Manpages & 940 & 2 & 35 & 22 & 78 & 82 & 75 & \textbf{87} \\ \cline{2-10} 
 & Bomber & 1036 & 2 & 4 & 14 & 84 & 79 & 78 & \textbf{90} \\ \cline{2-10} 
 & Learn Music Notes & 1352 & 4 & 19 & 30 & 62 & 62 & 64 & \textbf{75} \\ \cline{2-10} 
 & Dalvik Explorer & 5961 & 16 & 6 & 46 & 75 & 74 & 75 & \textbf{83} \\ \cline{2-10} 
 & Munch Life & 629 & 2 & 8 & 24 & 78 & 87 & 85 & \textbf{91} \\ \cline{2-10} 
 & K-9 Mail & 233860 & 27 & 365 & 188 & 9 & 7 & 8 & \textbf{13} \\ \cline{2-10} 
 & Yahtzee & 1997 & 2 & 16 & 16 & 59 & 52 & 71 & \textbf{76} \\ \cline{2-10} 
 & A2DP Volume & 13305 & 8 & 79 & 158 & 46 & 39 & 49 & \textbf{53} \\ \cline{2-10} 
 & ZooBorns & 2858 & 2 & 11 & 28 & 32 & 34 & 36 & \textbf{40} \\ \cline{2-10} 
 & Sanity & 21935 & 28 & 42 & 289 & 37 & 19 & 39 & \textbf{42} \\ \cline{2-10} 
 & Multi Sms & 2795 & 6 & 30 & 62 & 62 & 61 & 76 & \textbf{77} \\ \cline{2-10} 
 & Blokish & 6163 & 3 & 36 & 43 & 49 & 52 & \textbf{58} & \textbf{58} \\ \hline \hline
\multicolumn{2}{l|}{\textbf{Average}} & 21300 & 7 & 47 & 62 & 52 & 45 & 51 & \textbf{65} \\ \hline \hline
\multirow{21}{*}{\textbf{21}} & World Clock & 5560 & 2 & 16 & 19 & 25 & 95 & \textbf{98} & 97 \\ \cline{2-10} 
 & WordPress & 100829 & 63 & 1147 & 242 & 4 & 6 & \textbf{8} & 6 \\ \cline{2-10} 
 & aagtl & 48854 & 4 & 5 & 52 & 23 & 29 & \textbf{35} & 32 \\ \cline{2-10} 
 & Bites & 4354 & 5 & 24 & 52 & 42 & 35 & \textbf{57} & 53 \\ \cline{2-10} 
 & AutoAnswer & 471 & 1 & 21 & 3 & 18 & 9 & \textbf{25} & 18 \\ \cline{2-10} 
 & Alarm Klock & 9029 & 5 & 58 & 32 & 62 & 71 & \textbf{77} & 69 \\ \cline{2-10} 
 & Wikipedia & 233814 & 34 & 303 & 278 & 16 & 27 & \textbf{31} & 23 \\ \cline{2-10} 
 & myLock utilities & 2273 & 4 & 15 & 37 & 32 & 29 & \textbf{46} & 36 \\ \cline{2-10} 
 & Any Cut & 982 & 4 & 16 & 29 & 66 & 70 & \textbf{83} & 72 \\ \cline{2-10} 
 & Dialer2 & 2961 & 5 & 31 & 37 & 61 & 42 & \textbf{82} & 70 \\ \cline{2-10} 
 & PasswordMaker Pro & 17805 & 3 & 46 & 27 & 52 & 49 & \textbf{74} & 61 \\ \cline{2-10} 
 & aLogcat & 2751 & 2 & 13 & 29 & 59 & 72 & \textbf{80} & 66 \\ \cline{2-10} 
 & File Explorer & 13444 & 1 & 53 & 11 & 40 & 60 & \textbf{61} & 45 \\ \cline{2-10} 
 & Countdown Timer & 552 & 1 & 47 & 19 & 67 & 64 & \textbf{86} & 68 \\ \cline{2-10} 
 & Tomdroid & 22706 & 8 & 40 & 29 & 29 & 57 & \textbf{58} & 40 \\ \cline{2-10} 
 & Amaze & 64738 & 6 & 610 & 142 & 44 & 78 & \textbf{87} & 66 \\ \cline{2-10} 
 & NetCounter & 8059 & 3 & 33 & 47 & 44 & 68 & \textbf{79} & 53 \\ \cline{2-10} 
 & Random Music Player & 1730 & 4 & 18 & 15 & 55 & 59 & \textbf{88} & 58 \\ \cline{2-10} 
 & Import Contacts & 6435 & 4 & 68 & 18 & 27 & 42 & \textbf{79} & 37 \\ \cline{2-10} 
 & Jamendo & 13020 & 13 & 144 & 70 & 29 & 55 & \textbf{78} & 33 \\ \cline{2-10} 
 & Soundboard & 2296 & 2 & 103 & 23 & 28 & 54 & \textbf{100} & 33 \\ \hline \hline
\multicolumn{2}{l|}{\textbf{Average}} & 26793 & 8 & 134 & 58 & 39 & 51 & \textbf{67} & 49 \\ \hline \hline
\multirow{5}{*}{\textbf{5}} & aCal & 104542 & 25 & 249 & 96 & 14 & \textbf{29} & 26 & 22 \\ \cline{2-10} 
 & Addi & 75870 & 4 & 11 & 4 & 13 & \textbf{19} & 17 & 18 \\ \cline{2-10} 
 & Battery Dog & 2055 & 2 & 6 & 5 & 34 & \textbf{71} & 66 & 50 \\ \cline{2-10} 
 & Mileage & 44566 & 50 & 136 & 36 & 22 & \textbf{48} & 44 & 28 \\ \cline{2-10} 
 & Lock Pattern Generator & 1809 & 3 & 13 & 5 & \textbf{83} & \textbf{83} & 78 & 81 \\ \hline \hline
\multicolumn{2}{l|}{\textbf{Average}} & 45768 & 17 & 83 & 29 & 33 & \textbf{50} & 46 & 40 \\ \hline \hline
\textbf{1} & Book Catalogue & 38333 & 21 & 234 & 94 & \textbf{26} & 25 & 23 & 25 \\ \hline
\multirow{3}{*}{\textbf{3}} & Frozen Bubble & 22056 & 4 & 18 & 28 & 42 & - & 72 & 52 \\ \cline{2-10} 
 & aGrep & 3985 & 6 & 22 & 20 & 67 & - & 54 & 22 \\ \cline{2-10} 
 & Ringdroid & 12655 & 3 & 51 & 21 & 6 & 60 & - & 6 \\ \hline
\end{tabular}
\end{table*}

For the validity of comparison, we replicate the testing environment of Sapienz and Stoat for testing APEX and monkey. Additionally, 
we made our best effort to find test apps that have the same version as the ones used by both Sapienz and Stoat. The apps are selected based on the reported code coverage of monkey from the two publications. From the 93 apps tested by both Sapienz and Stoat, we selected 48 apps with which we can consistently achieve similar cover coverage running monkey.

To answer RQ2, we test APEX on the eight aforementioned
apps with a set of code targets as input, and examine
APEX's performance on targeted input generation.  Since
we used malware samples in this evaluation, our
motivation is to try to uncover possible malicious code
hiding deep in the source code. As such, we focus on
the ``hard-to-reach" targets, i.e., targets that
require more complex event sequences. In addition, we
also use two apps that have been used in the evaluation
of Collider for direct comparison. We also use a micro
benchmark Dragon to validate the capability of our
system. Dragon was developed to contain complex paths
that can be hard to reach. 

To identify code targets, we first conduct random
testing on the test apps using monkey with the same
setting as the previous study.  From monkey’s code
coverage reports, we identify the uncovered bytecode
blocks and use the first bytecode statements from these
blocks as code targets.  Considering that monkey is a
highly effective input generation
tool~\cite{ChoudharyASE15}, the code targets uncovered
by monkey is challenging enough.

\subsection{RQ1 - Code Coverage}
The overall code coverage on the 48 test apps 
are shown in Table~\ref{apex:table_cc}. The column
``\# LOC" shows the total lines of of dalvik 
bytecode in the test 
apps. The remaining columns show the code coverage 
result from Monkey(denoted as M), Sapienz(SA), 
Stoat(ST) and APEX(AP), respectively. 
The entry ``-" indicates that the tool does not have 
a code coverage result due to being unable to 
run the app.

Overall, APEX outperforms monkey and sapienz 
in 44 and 27 apps, respectively. Comparing to
Monkey's random fuzzing approach, and
Sapienz's genetic algorithm based on randomly
initialized event pool, APEX 
is able to expose more program behaviors by
systematically
building the fine-grained constraint aware 
GUI model. For example, when running the 
\textit{K9 Mail} app for the first time, 
user is required log in with an valid email 
address and password. The ``Next" button that 
leads to the next page remains disabled until 
the email address field matches a regular 
expression defined in the 
\textit{EmailAddressValidator} class. 
APEX can reliably proceed to the next page by: 
(1) identifying the constraint that guards 
the execution path to enable the ``Next” button, 
and (2) using the exact regex to generate a string 
that satisfies the constraint. 
In comparison, Monkey cannot reliably generate 
a valid email address since it fills text fields 
with random strings. Sapienz is capable of 
generating higher quality strings using string 
seeds collected from the app’s resource files, 
but it relies on Monkey to initialize the event 
pool, which makes it less effective in 
discovering GUI states beyond Monkey's coverage.

\begin{figure}[t]
\centering
	\includegraphics[width=0.4\textwidth]{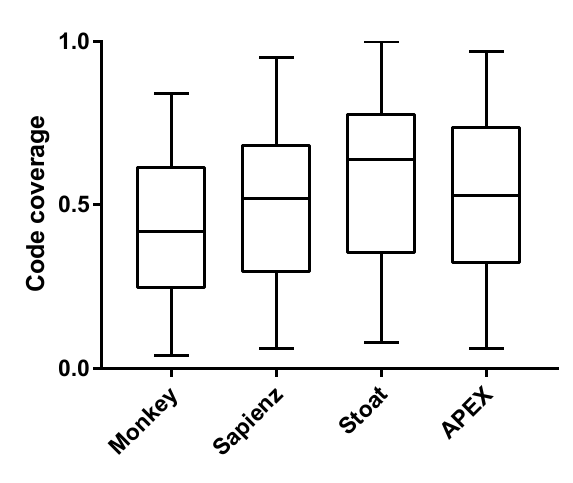}
	\caption{Code Coverage Distribution}
	\label{fig:apex_cc_boxplot}
\end{figure}

Stoat's two phase approach first generates a GUI model
using a relatively light-weight exploration strategy,
then uses Gibbs sampling to generate event sequences
from its GUI model.  Comparing to Stoat, APEX was able
to achieve higher code coverage in 18 of the 48 test
apps (listed as the first group of apps in
Table~\ref{apex:table_cc}).  Stoat was able to achieve
the highest coverage in 21 out of 48 apps (the second
group of apps).  Sapienz was able to acheive the
highest coverage in 5 apps (the third group of apps)
while Monkey is able to achieve the highest coverage in
1 app. Also note that there were three apps that
Sapienz or Stoat could not run. 

There were also 26 apps that APEX was unable to complete
the GUI exploration within the time limit.
Upon inspection, we learn that the 18 apps in the first
group generally have more GUI widgets and registered event
handlers than the other 26 apps, as shown in the three
rows named ``Average''. These results show the
different performance characteristics between Stoat and
APEX. Stoat's two phase approach first generates a GUI
model using a weighted exploration strategy, then uses
Gibbs sampling to generate event sequences from its GUI
model.  APEX focuses on building a richer GUI model for
deterministically exploring different program
behaviors.  In apps with less complicated GUI
structures, Stoat's GUI exploration can still generate
a good enough GUI model which enables the random
sampling phase to generate more effective event
sequences.  However, in the apps with more complex GUI
structures, Stoat might not be able to generate a good
enough GUI model in its first phase, while the
systematic exploration approach in APEX performs
better.

APEX shows better robustness by being able to run all
the test apps, while Stoat or Sapienz cannot run
three apps, e.g. \textit{aGrep}, \textit{Frozen
Bubble}, and \textit{Ringdroid}.  Overall, APEX
outperforms Monkey and Sapienz in majority of the apps,
while underperforming Stoat, as shown in
Figure~\ref{fig:apex_cc_boxplot}. Our results show that
APEX performs better than Stoat in apps with more
complex GUI structures while worse in apps with simple
GUI structures.  We believe that, in practice, APEX can
be used in complementary with the other tools such as
Stoat for higher code coverages, leading to better
testing outcomes.

\subsection{RQ2 - Targeted Input Generation} 

To test \textsc{APEx}'s effectiveness in generating
input for user-specific code targets, we test APEX on
eight apps with a set of ``hard-to-reach" targets
identified by the previously described process in the
testing methodology.

\begin{table}[h]
\centering
\caption{Target coverage of \textsc{APEx} on 8 selected apps}
\label{table:apex_tc}
	\begin{tabular}{l|r|r}
\hline
App Name      & \multicolumn{1}{l|}{Targets Reached} & \multicolumn{1}{l}{Max Sequence Length} \\ \hline
Dragon        & 5/5 (100\%)                          & 6                                       \\
Munchlife     & 20/29 (69\%)                         & 8                                       \\
TippyTipper   & 16/57 (28\%)                         & 5                                       \\
BattleStat    & 10/88 (11\%)                         & 7                                       \\
rLurker       & 12/141 (9\%)                         & 5                                       \\
AudioSidekick & 12/79 (15\%)                         & 4                                       \\
AWeather      & 4/170 (2\%)                          & 3                                       \\
Engologist    & 6/129 (4\%)                          & 3                                       \\ \hline
\end{tabular}
\end{table}

The result of APEX's target coverage is shown in 
Table~\ref{table:apex_tc}. The ``targets reached" 
column shows the number of reached targets over 
the total number of targets, the 
``Max Sequence Length" column shows the maximum 
length of event sequences generated for the targets.
Although the target coverage appears low in 
every app except for \emph{Dragon} and 
\emph{Munchlife}, it is also important to point 
out that the majority of the apps in this test 
group are sophisticated malwares designed with 
anti-analysis techniques. After inspecting the 
path constraints of the unsolved targets, the 
most common reason of failure is unsolved API
constraints. With our current signature based 
API constraint solver, we can only deal with a 
limited set of APIs. Therefore, APEX could not 
generate event sequences for the paths that 
involve API related constraints that it does 
not recognize.

\begin{table}[h]
\centering
\caption{Comparison in target coverage between Collider and APEX.}
	\begin{tabular}{l|r|r}
\hline
\multicolumn{1}{c|}{App Name} & \multicolumn{1}{c|}{\begin{tabular}[c]{@{}c@{}}Target coverage\\ by Collider\end{tabular}} & \multicolumn{1}{c}{\begin{tabular}[c]{@{}c@{}}Target coverage\\ by APEx\end{tabular}} \\ \hline
Tippytipper                   & 7/16 (44\%)                                                                                & 16/57 (28\%)                                                                          \\
Munchlife                     & 6/10 (60\%)                                                                                & 20/29 (69\%)                                                                          \\ \hline
\end{tabular}
\label{tab:apex_vs_collider}
\end{table}


Next we compare the targeted input generation result of
\textsc{APEx} with that of \emph{Collider}, a concolic
execution engine~\cite{collider}. In the evaluation of
\emph{Collider}, the target lines were selected from
unreached bytecode lines after running both
\emph{Monkey} and \emph{crawler}~\cite{crawler}, a GUI testing tool based on depth first exploration.
Since the source code of Collider is
not publicly available, and the specific code targets
are not specified from its publication, we followed
Collider's approach in generating code targets to make
the results comparable.

In \emph{TippyTipper}, \emph{Collider} was able to
reach 7 targets out of 16, while \textsc{APEx} has
reached 16 out of 57. In \emph{Munchlife},
\emph{Collider} was able to reach 6 out of 10 targets,
while \textsc{APEx} reached 20 out of 29. The
comparison in target coverage is shown in
Table~\ref{tab:apex_vs_collider}. We can see that
\textsc{APEx} has overall lower coverage rate while
reaching more targets than \emph{Collider}. Without a
thorough comparisons, it is difficult to determine
which tool performs better. However, \emph{Collider}'s
sequence generation requires a manually built GUI
model, while \textsc{APEx} does not make any
assumptions nor require manual effort with building the
GUI model. Overall, despite the problems and
limitations, \textsc{APEx} is easier to deploy than the
concolic execution engine \emph{Collider}, while able
to reach more targets in the same apps.

\subsection{Threats to Validity}
\label{eval_threats}

There are some limitations in APEX. First of all,
the inherent problem of path explosion threatens the
completeness of GUI traversal. Secondly, constantly
running background threads generate non-stop execution
logs that undermine the integrity of the symbolic execution.
For example, the music app ``Jamendo"
contains an ImageView widget to display album art. This
ImageView widget has a low level event handler
``onDraw()" that is repeatedly called as long as the
ImageView is being shown. This causes APEX to
misidentify the ``onDraw()" method as an implicitly
callback from another event, which results in a series
of incorrect path summaries. Thirdly, unsolved path
constraints might impede the GUI exploration towards
unexplored program states. In most of the
test app, unsolved path constraints involving system
APIs are the main cause for low code coverage.

We mitigate threats to validity by: 
(1) manually inspecting abnormally long callback sequences,
and refining the runtime monitor to identify auto-repeating
event handlers, (2) continuously adding API models to enhance
symbolic execution, which in turn reduce unsolvable constraint
for the constraint solver. For future work, we plan to use program
analysis techniques to systematically generate API models
for the symbolic execution.

%
%

\section{Related Work}

Prior to our development of {\sc APEx}, several input
generation tools that utilizes symbolic execution have
been proposed to serve various testing purposes. 

ACTEve~\cite{anand2012_acteve} is a GUI testing
framework that uses concolic execution instead of GUI
models to determine low level parameters of GUI events
such as the coordinates of tapping events. Symbolic
execution is also used to check whether an event's
impact of program state is relevant (read-only or
writes), and prune out event sequences that end in
irrelevant events. In terms of performance, ACTEve can
only generate event sequences with length no more than
four, which makes it less practical in testing modern
mobile applications.  Ganov et
al.~\cite{ganov2009_Barad} use symbolic execution to
generate an abstraction of the GUI interactions and
then generate concrete widget parameters to test Java
SWT applications.  AppIntent~\cite{yang2013_appintent}
performs symbolic execution selectively on a certain
set of event handlers to expose data leakage.
ConDroid~\cite{schutte2015_condroid} uses symbolic
execution and instrumentation to overwrite register
values during runtime to inspect specific program
behaviors. JPF-Android~\cite{van2012_JPFAndroid1}
,~\cite{van2014_JPFAndroid2} and
SymDroid~\cite{jeon2012_symdroid} are symbolic
execution engines specifically designed for Android
system. IntelliDroid~\cite{wong2016intellidroid} generates
targeted event sequences by solving method constraints on
the call graph path from the entry point to the target
method.
Our work is different than these work in that
we use concolic execution to systematically explore GUI
state transitions by leveraging the proposed
Constraint-Aware GUI Model.

The event sequence generation process of {\sc APEx} is
similar to Collider~\cite{collider}, an
Android input generation tool that uses concolic
execution and a GUI model to generate event sequences
for user-specified targets.  Given a specific program
code target, Collider uses symbolic execution to
identify a series of \emph{anchor events} that are
required to satisfy the path constraints of target code
execution path, then uses an existing GUI model to
generate event sequences that connects the anchor
events from the app entry to the final event. The
generated event sequences are then validated on a test
device. While Collider is capable of generating complex
event sequences for certain hard-to-reach targets. It
has the limitation of requiring an existing and sound
GUI model, which is not a trivial task especially
considering that Collider targets applications that
have complex GUI structures and control flows. Our work
overcomes this limitation by incrementally creating GUI
model as part of exploration.

Sig-Droid~\cite{mirzaei2015_sigdroid} is an input
generation framework that uses static analysis to
create two models of an application: \emph{behavior},
which exposes implicit calls among event handlers, and
\emph{interface}, which abstracts the GUI. Based on the
two models, Sig-Droid performs symbolic execution on
the event handler call graphs and then generates event
sequences based on the Interface Model.
TrimDroid~\cite{mirzaei2016reducing} also uses static
analysis to generate two models: the interface model
and the activity transition model. TrimDroid generates
event sequences based on extracting
GUI-induced dependencies using control-flow and data-flow
analysis. Similar to
{\sc APEx}, Sig-Droid and TrimDroid also aims to 
generate a more
detailed model of the application. However, one
limitation of Sig-Droid is that its Interface Model,
which is built by simple static analysis on XML files,
can suffer from incompleteness when dealing with
runtime generated GUI components, raising concerns
about validity and effectiveness of the event sequence
generation process. SimDroid on the other hand, does
not rely on the completeness of models and use the
extracted-dependencies to more precisely generate
event sequences and reduce combinatorics.

Model-based testing is also a commonly used testing
approach that focuses on using GUI models to generate test
cases.  MobiGuitar~\cite{amalfitano2015_mobiguitar} is
a dynamic GUI testing tool that builds a model of the
GUI by a depth-first exploration strategy.
PUMA~\cite{hao2014_puma} is a programmable GUI testing
framework that allows users to define specific actions
for available events during the exploration. Mahmood et
al.~\cite{mahmood2012_whitebox} proposes a cloud-based
testing framework that generates test cases 
from application model and executes them on multiple
emulators simultaneously.
A\textsuperscript{3}E~\cite{azim2013_a3e} is an input
generation tool that can explore the GUI using
depth-first exploration strategy and a more systematic
targeted exploration. Both exploration strategies of
A\textsuperscript{3}E are based on a statically
generated GUI model using taint analysis. 
DroidBox~\cite{li2017droidbot} is a lightweight GUI-model based
input generation tool that can generate a GUI model without
requiring source code or instrumentation. Comparing to
{\sc APEx}, a common limitation of these model-based
testing framework is their inability to generate event
sequences that can trigger specific program behaviors
beneath the GUI surface.
In terms of comparison among different testing
techniques, a recent study~\cite{generalframework}
proposed a unique platform that defines the
testing strategy and evaluates the effectiveness and
cost comparisons among existing testing tools.

Besides symbolic execution, various different
analysis techniques have been introduced to
improve the performance of input generation.
SwiftHand~\cite{choi2013_swifthand} uses machine
learning to learn the model of an application during
GUI exploration, while
EvoDroid~\cite{mahmood2014_evodroid} performs
evolutionary testing on Android apps.
EHBDroid~\cite{ehbdroid} is a unique GUI testing approach that
directly invokes the event handler methods through instrumentation,
as opposed to the traditional GUI based approach.
Sapienz~\cite{sapienz} is a search-based testing
approach that uses genetic algorithm 
and string seeding to generate event sequences.
Stoat~\cite{stoat} generates event sequences
using Gibbs sampling that favors events that
have higher probability to extend code coverage.

\section{Conclusion}
\label{conclusion}

One major challenge of testing GUI-based Android
applications is generating meaningful event sequences
that would allow software engineers or security
analysts to explore more application paths and/or
targets specific paths to exercise some desired
behaviors.  Modern approaches to tackle this issue tend
to generate input randomly or try to statically or
dynamically produce GUI models that help with input
generation. However, studies have shown that these two
approaches are not effective as the random input
generation approaches lack precision and the
model-based approaches lack completeness. 

In this work, we describe our implementation of
\emph{Android Path Explorer} ({\sc APEx}), an input
generation framework aiming to provide a systematic
exploration and event sequence generation for Android
applications.  We design {\sc APEx} to generate event
sequences that yield high code coverage as well as
event sequences that can target specific execution
paths. Unlike prior work, {\sc APEx} uses concolic
execution to (1) guide a systematic exploration of the
program behaviors to build an application model; and
(2) discover data dependencies between event handlers
and leverage the application model to generate concrete
event sequences. Our empirical evaluation using 48
applications shows that {\sc APEx} achieves higher
code coverage than monkey and sapienz in 44 and 27
of the test apps, respectively. While 
achieving lower code coverage than Stoat, we found
that {\sc APEx} performs better in the apps that
have more complex GUI structures.

We then conducted an evaluation to assess {\sc APEx}
ability to generate event sequences to reach specific
targets and found that it is moderately successful in
apps that do not rely too much on API and library
calls. In these apps, it can generate event sequences
that can reach 28\% to 100\% of selected targets. 

The major limitation of {\sc APEx} is that it can only
deal with a limited set of APIs due to its use of
signature based API constraint solver.  When an unknown
API is encountered, the solver cannot solve the
constraint, and prematurely terminates the event
sequence generation process. Effectively dealing with
these system APIs and libraries is still a great
challenge and continues to be the focus of
\textsc{APEx} development. As we continue to refine its
implementation, we plan to contribute to our software
engineering research community by making {\sc APEx}
publicly available for other researchers to use.

\bibliographystyle{abbrv}
\bibliography{References}

\end{document}